\documentclass{jnmp}
\usepackage{amsmath}

\def\prof{\vrule depth 3.5pt width 0pt}
\def\sti#1#2#3{\prof{\smash{\hbox{\vtop{\hbox{$#1$}%
\hbox{\raise#2pt\hbox{$\mkern#3mu\tilde{}$}}}}}}}
\def\wdt{\sti W 6 8}
\def\wmdt{\sti w 6 6}
\def\kwdt{\sti {\cal W} 8 8}
\def\mwdt{\sti {\mathtt{w}} 8 6}

\JNMPnumberwithin{equation}{section}

\begin{document}

\renewcommand{\evenhead}{B~Grammaticos, A~Ramani and Y~Ohta}
\renewcommand{\oddhead}{A Unified Description of the
Asymmetric $q$-P$_{\rm V}$
and {\rm d}-P$_{\rm IV}$ Equations}

\thispagestyle{empty}

\FirstPageHead{10}{2}{2003}{\pageref{Grammaticos-firstpage}--\pageref{Grammaticos-lastpage}}{Article}

\copyrightnote{2003}{B~Grammaticos, A~Ramani and Y~Ohta}

\Name{A Unified Description of the
Asymmetric \\
$\boldsymbol{q}$-P$\boldsymbol{{}_{\rm V}}$
and d-P$\boldsymbol{{}_{\rm IV}}$ Equations\\
 and their Schlesinger Transformations}
\label{Grammaticos-firstpage}

\Author{B~GRAMMATICOS~$^{\dag^1}$, A~RAMANI~$^{\dag^2}$ and Y~OHTA~$^{\dag^3}$}

\Address{$^{\dag^1}$~GMPIB, Universit\'e Paris VII, Tour 24-14, 5$^e$\'etage, case 7021,
75251 Paris, France \\[10pt]
$^{\dag^2}$~CPT, Ecole Polytechnique, CNRS, UMR 7644, 91128 Palaiseau, France \\[10pt]
$^{\dag^3}$~Department of Applied Mathematics,
Faculty of Engineering, Hiroshima University,\\
$\phantom{^{\dag^3}}$~1-4-1 Kagamiyama, Higashi-Hiroshima 739-8527, Japan}

\Date{Received July 3, 2002; Revised September 27, 2002;
Accepted October 1, 2002}

\begin{abstract}
\noindent
We present a geometric description, based on the affine Weyl group
E$_6^{(1)}$, of two discrete analogues of the
Painlev\'e VI equation, known as the asymmetric $q$-P$_{\rm V}$ and
asymmetric d-P$_{\rm IV}$. This approach allows
us to describe in a unified way the evolution of the mapping along
the independent variable and along the various
parameters (the latter evolution being the one induced by the
Schlesinger transformations). It turns out that both
discrete Painlev\'e equations exhibit the property of self-duality:
the same equation governs the evolution along
any direction in the space of E$_6^{(1)}$.
\end{abstract}

\section{Introduction }
The study of integrable discrete systems has revealed the most
interesting fact that these systems present many
common points with their continuous counterparts. One of these points
was the role played by singularities in
discrete integrability~[1]. By examining the singularities of a given
mapping, and requesting that those which appear
spontaneously do not propagate {\it ad infinitum}, we were able to
derive the discrete analogue of the Painlev\'e
equations~[2].

Painlev\'e equations were introduced one century ago in order to
extend to the nonlinear domain the notion of special
function defined by a differential equation.  The Painlev\'e
transcendents were discovered in that way. Discrete
forms of the Painlev\'e equations were discovered as soon as 1939~[3]
(and are present in essence if not in precise
form in the work of Laguerre~[4] which precedes that of Painlev\'e)
but were not recognised as such till recently.
However it was only after the discovery of the singularity
confinement property that  the study of the discrete
Painlev\'e equations received a substantial boost. The principle for
their derivation is simple: start from an
integrable autonomous mapping (typically one of the QRT~[5] family)
which contains free parameters and apply the
singularity confinement criterion in order to fix the~$n$-dependence
of the parameters. This approach made possible
the derivation of the $q$-analogues of the Painlev\'e equations. The
latter are mappings where the independent
variable enters not in an additive but, rather, in a multiplicative
way. The first instance of such an equation was~[6]:
\begin{equation}
\overline x \underline x
=\frac{cd(x-a\lambda^n)(x-b\lambda^n)}{(x-c)(x-d)},
\end{equation}
where $x=x(n)$, $\overline x=x(n+1)$, $\underline x=x(n-1)$ and $a$,
$b$, $c$, $d$ are constant.
This equation was shown to be a $q$-discrete analogue of P$_{\rm
III}$ provided one discards the possible even-odd
dependence of the coefficients. However when the full freedom of the
coefficients is restored and one rewrites the
equation in an asymmetric form (the term `asymmetric' being used here
in the QRT sense):
\begin{subequations}
\begin{gather}
y \underline y =\frac{cd(x-a\lambda^n)(x-b\lambda^n)}{(x-p)(x-q)},\\
\overline x  x =\frac{pq(y-r\lambda^n)(y-s\lambda^n)}{(y-c)(y-d)},
\end{gather}
\end{subequations}
where $a$, $b$, $c$,
$d$, $p$, $q$, $r$, $s$ are constants constrained by $pqrs=\lambda abcd$.
It can be shown, as was done by Jimbo and Sakai~[7], that this
equation is a discrete form of P$_{\rm VI}$. The
interesting property of equations where no artificial limitation of
the richness of their parameters is imposed is
the self-duality, first discovered in~[8]. While studying the action
of Schlesinger transformations of
discrete Painlev\'e equations it was found that the same equation
governs the evolution along the independent variable
and the Schlesinger-induced shifts of parameters. We know today that
not all discrete Painlev\'e equations posses the
property of self-duality~[9], however it was this discovery which
made possible a geometrical description~[10] of the
discrete Painlev\'e equations and their classification. This
geometrical description relies heavily on affine Weyl
groups (as the one of the continuous Painlev\'e equations) and was
dubbed in~[11] the ``Grand Scheme''.

In this paper we shall present the geometrical, affine Weyl
group-based, description of two equations. The first
is known as asymmetric d-P$_{\rm IV}$:
\begin{gather}
({\overline x}+y)(y+x)=\frac{(y-a)(y-b)(y-c)(y-d)}{(y-z-\kappa/2)^2-e^2},\nonumber\\
(y+x)(x+{\underline y})=\frac{(x+a)(x+b)(x+c)(x+d)}{(x-z)^2-f^2},
\end{gather}
where $z=\kappa n+\mu$ and the constants $a$,
$b$, $c$, $d$, $e$, $f$ satisfy the
constraint $a+b+c+d=0$.
The second is known as the asymmetric $q$-P$_{\rm V}$
\begin{gather}
({\overline x}
y-1)(yx-1)=\frac{rs\lambda^{2n+1}(y-a)(y-b)(y-c)(y-d)}{(y-p\lambda^n)(y-q\lambda^n)},
\nonumber\\
(yx_-1)(x{\underline y}-1)=\frac{pq\lambda^{2n-1}(x-1/a)(x-1/b)(x-1/c)(x-1/d)}
{(x-r\lambda^n)(x-s\lambda^n)},
\end{gather}
where the constants $a$, $b$, $c$, $d$, $p$, $q$, $r$, $s$ satisfy the constraint
$pq=\lambda abcdrs$.
These equations have been first proposed in~[12] and further
studied in [13, 14] but their geometrical description had not been
presented yet.

\section{The geometry of the E$\boldsymbol{{}_6^{(1)}}$ weight lattice}

\vspace{-2mm}

Our basic assumption is that the $\tau$-functions of both the
discrete Painlev\'e equations under study live on the
points of the weight lattice of the affine Weyl group E$_6^{(1)}$.
It turns out that  there is no orthonormal basis
invariant under the action of the group. In analogy to what we did in
the case of E$_7^{(1)}$~[15] we will choose an
orthogonal basis where all vectors are not equivalent. The squared
length of the first vector will be chosen equal to
1/2, while that of the five others will be taken equal to  3/2.
In this basis, these points are such that their coordinates
are of the form $(a;b_1,b_2,b_3,b_4,b_5)$, where $a$ (the coordinate
along the squared-length-1/2 vector) {\it and}
the coordinates $b_i$'s (on the five squared-length-3/2 vectors)
are either {\it all} integer or {\it  all}
half-integers, with the additional constraint that the sum of all
coordinates ($a$ and $b$'s) is even.

The origin satisfies these requirements. It has 54 nearest-neighbours
(NN) of the following form. First, $(\pm
2;0,0,0,0,0)$, then 20 such that $a=\pm 1$ and {\it one} nonzero
coordinate $b_i=\pm1$ while the other four vanish
and finally 32 where both $a$ and each $b_i$ have absolute value 1/2
and arbitrary signs, up to the constraint that
the total number of minus signs be odd, to ensure that the total sum
be even, thus leading to 32 rather than 64
points. The squared distance of each of these points to the origin is
always 2, be it 4/2, $1/2+3/2$ or $1/8+5(3/8)$.
Though in this particular basis these points look very different, they are in
fact all  equivalent. They
define 27 directions along which NV's vectors exist and this notation
stands for `Nearest-neighbour-connecting
Vectors'. Contrary to the case of E$_7^{(1)}$~[15] where we could not
fix consistently the  orientations of the NV's,
for the case of E$_6^{(1)}$ if we choose as our oriented NV's
$(2;0,0,0,0,0)$, $(-1;0,\dots,
\pm1,\dots)$ and
$(1/2;\pm 1/2,\pm 1/2,\pm 1/2,\pm 1/2,\pm 1/2)$
(with odd number of
minus signs in the last case), then the scalar products of any two
distinct NV's is either $-1$ or 1/2 but never 1 nor
$-1/2$. Moreover each NV has scalar product $-1$ with exactly 10 NV's
and 1/2 with the 16 others. This shows again
that all NV's are equivalent. Also their total sum is zero, as can be
easily checked. (For instance see that the
scalar product of the sum with any NV is $2+10(-1)+16(1/2)=0$).

The sum of two NV's of scalar product $-1$ has squared length~2, and
in fact is just the opposite of some other NV. So
to get further away from the origin, to a $\tau$ which is
next-nearest-neighbour (NNN) of the origin, we have to
translate by the sum of two vectors of  scalar product
$-1/2$, i.e. the {\it difference} of two NV's of scalar product~1/2,
to get an NNV of squared length~3. Here NNV stands
for `Next-Nearest-neighbour-connecting Vectors'. Contrary to the NV's
that can be consistently oriented, the NNV's
cannot (the orientation of the NV's does not carry over because we
are taking differences). Though there are
216~$(=27 \times 16/2$) pairs of NV's of scalar product 1/2, there are only
36 NNV's, since each can be obtained in six
different ways. For instance the NNV  $(0;1,1,0,0,0)$ is the difference
of the pairs
$\{(-1;1,0,0,0,0), (-1;0,-1,0,0,0)\}$ and
$\{(-1;0,1,0,0,0), (-1;-1,0,0,0,0)\}$ but also of four pairs
$\{(1/2;1/2,1/2,\pm 1/2,\pm 1/2,\pm 1/2),
(1/2;-1/2,-1/2,\pm 1/2,\pm 1/2,$ \linebreak $\pm 1/2)\}$
where the three signs are the same for
two vectors in a pair, and are constrained by the even-sum rule to
comprise an odd number of minus signs. In the
basis we consider there are 20 NNV's of this form, with  $a$ and
three $b_i$'s vanishing and the two others of
absolute value 1 with arbitrary sign, (but only 20 rather than
$40=4(5\times 4/2)$ because we ignore the orientation
of the NNV's) and 16 more which have $a=-3/2$ and the same values of
the $b_i$'s as the 16 last NV's, but again, all
NNV's are equivalent.

\section{The nonlinear variables and the Hirota--Miwa equation}

Consider a 2-dimensional plane containing the origin, say, and two
$\tau$'s, both NN's of the origin, such
that the relevant NV's have scalar product 1/2, for instance
$(-1;1,0,0,0,0)$ and $(-1;0,-1,0,0,0)$. They are in
NNN position with respect to each other, since their squared-distance
is~3. But what is
interesting is to consider the fourth point in the parallelogram, the
one obtained in translating the origin by the sum
of these two NV's. This point (in our case $(-2;1,-1,0,0,0)$) is at a
squared-distance~5 from the origin,  and is in
next-next-nearest-neighbour (NNNN) position with respect to it. This means
that the center of our parallelogram, midpoint of a pair of  NNN
$\tau$'s is {\it also} the midpoint of pair of two
NNNN
$\tau$'s. Moreover, since there are six different ways to write an
NNV as the difference of two NV's, the same point
is altogether the midpoint of six different pairs of NNNN $\tau$'s.
In our case, the point $X$,
$(-1;1/2,-1/2,0,0,0)$, midpoint of the NNN pair
$\{(-1;1,0,0,0,0),(-1;0,-1,0,0,0)\}$, is also  midpoint of the NNNN
pairs, not only
$\{(0;0,0,0,0,0),(-2;1,-1,0,0,0)\}$ but also
$\{(-2;0,0,0,0,0),(0;1,-1,0,0,0)\}$ and four pairs of the form
$\{(-1/2;1/2,-1/2,\pm 1/2,\pm 1/2,\pm1/2),
(-3/2;1/2,-1/2,\mp 1/2,\mp 1/2,$ \linebreak $\mp 1/2)\}$,
where the three last signs have opposite
values in the two points of a given pair, the number of minus signs
being odd (resp.\ even) for the first (resp.\
second) point in each pair, which guarantees that the even-sum rule
always holds. These points, midpoints of one pair
of NNN $\tau$'s and  of six pairs of NNNN $\tau$'s, are the points
where we will define nonlinear variables, $X$
or~$x$ for the asymmetric d-P$_{\rm IV}$ and $q$-P$_{\rm V}$ equations
respectively.

Note that, contrary to the NNV's which cannot be oriented
consistently, the NNNV's are sums of NV's and can all be
consistently oriented by carrying over the orientation of the NV's.
The six NNNV's around the site of a particular
nonlinear variable are not independent: they all lie in the same
hyperplane orthogonal to the NNV joining the pair of
NNN $\tau$'s around the same site. In fact one can easily convince
oneself that their sum vanishes: each of them has
scalar product $-1$ with each of the five others (we recall that
their squared sum is precisely~5). In the particular
case we are considering, the correctly oriented NNNV's are
$(-2;1,-1,0,0,0)$, $(-2;-1,1,0,0,0)$, and four of the form
$(1;0,0,\pm 1,\pm 1,\pm 1)$ with an odd number of minus signs (such
a vector, with an even number of minus signs,
would still be a valid NNNV between {\it some} $\tau$'s, as we shall
see later, but {\it not} around the point we are
considering). Let us choose  some point $O'$, not necessarily the
origin of coordinates, and call $C_i$ the
scalar products of the vector $\overrightarrow {O'X} $ with these six
NNNV's, and introduce
$c_i=q^{C_i}$, for some number $q$. We have of course $\sum_i C_i=0$
and $\prod_i c_i=1$.

We are now in a position to express the value of the nonlinear
variable at the point we considered. Let $\psi$ be the
product of the two NNN $\tau$'s (in our case,
$\tau_x$ at $(-1;1,0,0,0,0)$ and $\tau'_x$ $(-1;0,-1,0,0,0)$) and
$\phi_i$ the product of two NNNN
$\tau$'s at the ends of the vector that defines  $C_i$. Then, for
asymmetric d-P$_{\rm IV}$ we have:
\begin{equation}
X=C_i+\frac{\phi_i}{ \psi}
\end{equation}
 and  for asymmetric $q$-P$_{\rm V}$
\begin{equation}
x=c_i+c_i^{1/3}\frac{\phi_i}{\psi}.
\end{equation}
 This  implies
compatibility conditions, which are non-autonomous
Hirota--Miwa~[16] systems:
\begin{gather}
\phi_i-\phi_j+(C_i-C_j) \psi=0,\\
c_i^{1/3}\phi_i-c_j^{1/3}\phi_j+(c_i-c_j) \psi=0.
\end{gather}
The
set of equations (3.3) (resp.  (3.4)) around {\it
all} possible sites for nonlinear variables  is overdetermined but
consistent over the entire lattice and is nothing
but the bilinear form of the asymmetric d-P$_{\rm IV}$ (resp.
$q$-P$_{\rm V}$) equation.

\section{The nonlinear equations}

Around each site like $X$, among the 27 NV's, exactly 12  are used up
in constructing, in pairs, the 6
NNNV's around $X$, (or, equivalently, lead by their differences to
the NNV around $X$). There are 15 NV's left. On
the other hand there are exactly 15 ways to choose two among the six
NNNV's. It turns out that, for any choice of a
pair of NNNV around $X$, the sum of these vectors is exactly twice
the opposite of one of the 15 remaining NV's. Not
only that, but if one translates
$X$ by half of any of these NV's in either direction, one finds another
point $Y$ where a nonlinear variable can be
defined. This was by no means obvious: if we translate $X$ by half of
any of the first 12 NV's, the resulting point
would not be the midpoint of two $\tau$'s in NNN position. To be
specific, we easily see that no pair of NV's
containing $(2;0,0,0,0,0)$  allows to construct, by difference, the NNV
$(0;1,1,0,0,0)$ around $X$. Conversely, if we
take $(-2;1,-1,0,0,0)$, $(-2;-1,1,0,0,0)$ among the 6 NNNV's around
$X$, their sum is twice the opposite of this NV.
Thus the point $X$ can be translated by $(\pm 1;0,0,0,0,0)$ to lead to
new sites $Y$ $(0; 1/2,- 1/2,0,0,0)$ and
${\underline Y}$  $(-2; 1/2,- 1/2,0,0,0)$, where nonlinear variables can
be constructed. Note that the environments of
$Y$, ${\underline Y}$ in terms of $\tau$'s are identical (since the distance
between these two points is a full NV) but are
{\it not} the same as that of $X$. For instance the NNV around
$Y$ (and ${\underline Y}$) is not   $(0;1,1,0,0,0)$ but $(0;-1,1,0,0,0)$, and the
NNNV's also differ from those at $X$, being
$(-2;1,1,0,0,0)$, $(-2;-1,-1,0,0,0)$ and of the form
$(1;0,0,\pm 1,\pm 1,\pm 1)$ but with an {\it even} number of
minus signs. So, if ($a;b_i$) are the components of the  vector
$\overrightarrow{O'X}$ the six $C_i$'s around $X$
come into two groups (this is because we are distinguishing a
specific NV, that of the direction $XY$; from an
absolute point of view all $C_i$'s are equivalent):
\begin{gather}
-a+3/2(b_1-b_2)\equiv -2Z+p,\nonumber\\
-a-3/2(b_1-b_2)\equiv -2Z-p
\end{gather}
 on the one hand,
\begin{gather}
a/2-3/2(b_3+b_4+b_5)\equiv Z+\alpha,\nonumber\\
a/2+3/2(-b_3+b_4+b_5)\equiv Z+\beta,\nonumber\\
a/2+3/2(b_3-b_4+b_5)\equiv Z+\gamma,\nonumber\\
a/2+3/2(b_3+b_4-b_5)\equiv Z+\delta
\end{gather}
 on the other, where $Z=a/2$
and obvious notations for $p$, $\alpha$,
$\beta$, $\gamma$ and
$\delta$, with $\alpha+\beta+\gamma+\delta=0$. The six $D_j$'s around $Y$ are
$-(a+1)+3/2(b_1+b_2)\equiv (-2Z-1+r)$, $-(a+1)-3/2(b_1+b_2)\equiv
(-2Z-1-r)$ and four of the form
$Z+1/2-\alpha$, etc., the ones around ${\underline Y}$ being the same up to
replacing $Z+1/2$ by $Z-1/2$.

Consider now  one of the pairs of NNNN $\tau$'s around $X$ associated
to one of the four last $C_i$'s, say
$Z+\alpha$. The NNNV is $(1;0,0,1,1,1)$ and the two relevant $\tau$'s are
$\tau_\alpha$ at
$(- 1/2; 1/2,- 1/2,- 1/2,-1/2,-1/2)$ and
$\tau'_\alpha$ at
$(-3/2;1/2,-1/2,1/2,1/2,1/2)$. The
first one  also forms an NNNN pair near $Y$ while the
second forms an NNNN pair near
${\underline Y}$, both corresponding to the $D_j$ involving $\alpha$. Thus if
for instance $\tau_\alpha$ vanishes, we know the
values of both $X$ and $Y$ (from (3.1) and (3.2) because the
corresponding $\phi$ vanishes). At this point we must start
to separate the study of
$q$-P$_{\rm V}$ from that of d-P$_{\rm IV}$.

Let us first consider d-P$_{\rm IV}$. Then, if $\tau_\alpha$
vanishes, from (3.1) we have $X=Z+\alpha$  and from its
analogue at $Y$, $Y=Z+1/2-\alpha$, because the quantity $\phi$ in that case is
$\tau_\alpha\tau'_\alpha$ for $X$ and
$\tau_\alpha\overline\tau'_\alpha$ for $Y$ where
$\overline\tau'_\alpha$ is at the point
$(1/2;1/2,-1/2,1/2,1/2,1/2)$
(translated from the site of $\tau'_\alpha$ by the full NV
along
$XY$). The quantities at the denominators of $X$ and $Y$ are
respectively $\tau_x\tau'_x$  and $\tau_0\tau_y$ where
we recall that $\tau_x$ is at point $(-1;1,0,0,0,0)$ and $\tau'_x$ at
$(-1;0,-1,0,0,0)$, while
$\tau_0$ is at the origin $(0;0,0,0,0,0)$ and $\tau_y$ at
$(0;1,-1,0,0,0)$ symmetric of the origin with respect to
$Y$. Consider now the quantity  $X+Y-2Z-1/2$. Computing $X$  through
(3.1) and the appropriate instance of (4.2) (namely the first one),
and similarly
$Y$ through  their  analogues for we have
\begin{gather}
X+Y-2Z-1/2=\frac{\tau_\alpha\tau'_\alpha}{\tau_x\tau'_x}
+\frac{\tau_\alpha\overline\tau'_\alpha}{\tau_0\tau_y}
=\frac{\tau_\alpha(\tau_0\tau_y\tau'_\alpha+\tau_x\tau'_x\overline\tau'_\alpha)}
{\tau_x\tau'_x\tau_0\tau_y}.
\end{gather}  The four numbers $\alpha$,
$\beta$, $\gamma$ and $\delta$ are
equivalent, and to each of them one can assign a $\tau$ which forms
an NNNN pair around both $X$ and $Y$. Whenever
each of them vanishes, both $X$ and
$Y$ take the corresponding $C_i$ and $D_j$ value, the sum of which is
$2Z+1/2$ in all four cases. This quantity is
the scalar product of the NV $(2;0,0,0,0,0)$ in the direction of $XY$
with the vector from $O'$ to the midpoint of
$XY$. The left hand side of (4.3) thus vanishes whenever one of these four
$\tau$'s vanish. It follows that the numerator of the right hand side
must also vanish, since it is equal to the
product of the vanishing quantity
$X+Y-2Z-1/2$ by a product of $\tau$'s which, being entire functions,
may  never become infinite. Hence this numerator
is proportional not only to
$\tau_\alpha$ but also to the product of the three others. It follows that:
\begin{equation}
\tau_0\tau_y\tau'_\alpha+\tau_x\tau'_x\overline\tau'_\alpha=K\tau_\beta\tau_\gamma\tau_\delta .
\end{equation}
 By
homogeneity, $K$ does not depend on any other $\tau$'s. This is thus
a trilinear equation satisfied by the
$\tau$'s. One does not need to impose it independently of (3.3). In
fact one can show, eliminating repeatedly various $\tau$ functions
between (3.3)
and its analogues at appropriate points, that (4.4) is just a
consequence of the
bilinear equation (3.3) and moreover that $K$ is
just $-1$. So  one gets:
\begin{equation}
\tau_0\tau_y\tau'_\alpha+\tau_x\tau'_x\overline\tau'_\alpha
+\tau_\beta\tau_\gamma\tau_\delta=0 .
\end{equation}
 Indeed
the three products play the same role, each being the product of
three $\tau$ forming an equilateral triangle the
side of which has squared-length 3 (indeed, each side is an NNV), all
three triangles having the same center of
mass, namely the point $G$ at
$(-1/2;1/2,-1/2,1/6,1/6,1/6)$. These
nine $\tau$'s are the
only ones at distance 1 from $G$, and there is no other way to
arrange them in three such equilateral
triangles. Equation (4.5) is thus an instance of a very general
trilinear equation which is true around every point on
the lattice equivalent to $G$, as a consequence of (3.1). The numerator of the
r.h.s. of (4.3)  can be replaced by a monomial using (4.5) and we find:
\begin{equation}
X+Y-2Z-1/2=-\frac{\tau_\alpha\tau_\beta\tau_\gamma\tau_\delta}
{\tau_x\tau'_x\tau_0\tau_y}.
\end{equation}

Clearly the
same reasoning can be done for $X$ and ${\underline Y}$. If  $\tau'_\alpha$ at
$(-3/2;1/2,-1/2,1/2,1/2,$ $1/2)$ vanishes,
$X$ has still the value $Z+\alpha$, but while we
have no direct information on $Y$, the value of ${\underline Y}$ is
$Z-1/2-\alpha$, because $\tau'_\alpha$ does form an NNNN
pair around ${\underline Y}$. Thus one has $X+{\underline Y}=2Z-1/2$, but this is also
true when the analogous $\tau'_\beta$, etc.
vanish. So again:
\begin{equation}
X+{\underline Y}-2Z+1/2=-\frac{\tau'_\alpha\tau'_\beta\tau'_\gamma\tau'_\delta}
{\tau_x\tau'_x\underline\tau_0\underline
\tau_y},
\end{equation}   where $\underline\tau_0\underline\tau_y$ is the
$\psi$ corresponding to ${\underline Y}$, at points
$(-2;0,0,0,0,0)$ and $(-2;1,-1,0,0,0)$ Taking the product we find:
\begin{equation}
(X+Y-2Z-1/2)(X+{\underline Y}-2Z+1/2)=
\frac{\tau_\alpha\tau_\beta\tau_\gamma\tau_\delta}{\tau_x\tau'_x\tau_0\tau_y}
\frac{\tau'_\alpha\tau'_\beta\tau'_\gamma\tau'_\delta}
{r\tau_x\tau'_x\underline\tau_0\underline
\tau_y}.
\end{equation}
 From (3.1) and (4.2) we recognize at the
numerator of the right-hand side the product of the
numerators of the quantities ($X-Z-\alpha$), etc.\ thus:
\begin{gather}
(X+Y-2Z-1/2)(X+{\underline Y}-2Z+1/2)\nonumber\\
\qquad {}=
(X-Z-\alpha)(X-Z-\beta)(X-Z-\gamma)(X-Z-\delta)
\frac{\tau_x^2{\tau'_x}^2}{\tau_0\tau_y\underline\tau_0\underline\tau_y}.
\end{gather}

But $\tau_0$ and
$\underline\tau_y$, on the one hand, and $\underline\tau_0$ and
$\tau_y$, on the other, are precisely the last two
pairs of NNNN $\tau$'s around $X$ and from (3.1) and (4.1), the last
factor in (4.9) is just the inverse of
$(X+2Z-p)(X+2Z+p)$. Thus:
\begin{gather}
(X+Y-2Z-1/2)(X+{\underline Y}-2Z+1/2)\nonumber\\
\qquad {}=
\frac{(X-Z-\alpha)(X-Z-\beta)(X-Z-\gamma)(X-Z-\delta)}{(X+2Z-p)(X+2Z+p)}.
\end{gather}
This is one of the two equations of the system defining the
asymmetric d-P$_{\rm IV}$ equation~[14], though in a
slightly unusual form. In order to obtain the other one, we need to
consider the point $\overline X$ at
$(1;1/2,-1/2,0,0,0)$. The couple of points ($Y,\overline
X$) is translated from (${\underline Y},X$) by exactly
one NV, so the environment is the same and from (4.7) one gets (with
obvious notations)
\begin{gather}
\overline
X+Y-2Z-3/2=-\frac{\overline\tau'_\alpha\overline\tau'_\beta\overline\tau'_\gamma
\overline\tau'_\delta}{\overline\tau_x\overline\tau'_x\tau_0\tau_y}.
\end{gather}

Multipling with (4.6) we see
appearing on the right hand side products of $\tau$'s near $Y$.
Finally what we get is the analogue of (4.10):
\begin{gather}
(X+Y-2Z-1/2)(\overline X+Y-2Z-3/2)\\
\qquad {}=\frac{(Y-Z-1/2+\alpha)(Y-Z-1/2+\beta)(Y-Z-1/2+\gamma)
(Y-Z-1/2+\delta)}{(Y+2Z+1-r)(Y+2Z+1+r)}.\nonumber
\end{gather} To
recover the usual form, we redefine $X$,
$Y$, ${\underline Y}$ and $\overline X$ by adding to them the relevant value of
the independent variable ($Z$, $Z+1/2$, $Z-1/2$
and $Z+1$, respectively) to get in the translated variables:
\begin{subequations}
\begin{gather}
(X+Y)(X+{\underline Y})=\frac{(X-\alpha)(X-\beta)(X-\gamma)(X-\delta)}{(X+3Z-p)(X+3Z+p)},\\
(X+Y)(\overline X+Y)=\frac{(Y+\alpha)(Y+\beta)(Y+\gamma)
(Y+\delta)}{(Y+3Z+3/2-r)(Y+3Z+3/2+r)}.
\end{gather}
\end{subequations}
 Equation
(4.13) is exactly the asymmetric  d-P$_{\rm IV}$ equation [14] up to
a redefinition of the variable $Z$.

The case of the asymmetric  $q$-P$_{\rm V}$ equation is similar but
slightly more complicated. The positions of the
relevant $\tau$'s are exactly the same, so we will keep the same
names, but the equations coming from (3.2) are not
quite the same. Let us follow the corresponding steps. When
$\tau_\alpha$ vanishes, we have $x=q^{Z+\alpha}$,
$y=q^{Z+1/2-\alpha}$ so
$xy=q^{2Z+1/2}$. Let us compute in all generality, the quantity
$xyq^{-2Z-1/2}$. To get  $x$ we use (3.2) with the instance of (4.2)
involving $\alpha$, and similarly for $y$.  We find
\begin{gather}
{xy q^{-2Z-1/2}}=1+q^{-2(Z+\alpha)/3}\frac{\tau_\alpha\tau'_\alpha}{\tau_x\tau'_x}
\nonumber\\
\qquad {}+q^{-2(Z+1/2-\alpha)/3}\frac{\tau_\alpha\overline\tau'_\alpha}{\tau_0\tau_y}
+q^{-(4Z+1)/3}\frac{\tau_\alpha^2\tau'_\alpha\overline\tau'_\alpha}
{\tau_x\tau'_x\tau_0\tau_y}
\end{gather}
so indeed $\left(xyq^{-2Z-1/2}-1\right)$  vanishes when $\tau_\alpha$ does. But though
we have {\it expressed} this quantity with emphasis on $\alpha$, the
three other quantities $\beta$,
$\gamma$ and $\delta$ play the same role and $\left(xyq^{-2Z-1/2}-1\right)$ also
vanishes whenever the associated $\tau$  does.
With the same argument as above, it follows that the numerator of the
right hand side of (4.14), after subtracting~1,
must be proportional to the product of these $\tau$'s. So we get the
analogue of equation~(4.4)
\begin{gather}
q^{-2(Z+\alpha)/3}{\tau'_\alpha\tau_0\tau_y}
+q^{-2(Z+1/2-\alpha)/3}{\overline\tau'_\alpha\tau_x\tau'_x}
+q^{-(4Z+1)/3}\tau_\alpha\tau'_\alpha\overline\tau'_\alpha=K\tau_\beta\tau_\gamma\tau_\delta.
\end{gather}
As in the case of (4.4), homogeneity shows that $K$ does not depend
on any $\tau$'s. But contrary to the previous
case, $K$ is not a constant, but a function of the point on the
lattice. Let us look at this trilinear equation more
closely. Three of the products are the same as in (4.5), but they
have a prefactor which is not unity. The last
product on the left-hand-side was not present in (4.4). It involves
one new $\tau$, namely $\tau_\alpha$, plus one
$\tau$ from two of the three other products. These three $\tau$'s
form an isosceles triangle of sides of
squared-length 2, 5 and 5. Its center of mass is still the same point
$G$ as that of the three others, but now
$\tau_\alpha$ is at squared-distance 2 from it. There are thus many
such equations that are true around the same
point, one for each of the 27 ways to pick one $\tau$ out of two of
the three three-$\tau$'s products in (4.5), and
complete the triangle of center of mass $G$ to a
$\tau$  at squared-distance 2 from it.

In order to obtain the value of
$K$, one repeatedly applies the analogue of (3.4) around various
points. One can actually prove (4.15) and obtain the
value of $K$ which is \linebreak $-q^{-(8Z+2)/3}$. Rewriting (4.15), divided by ($-K$):
\begin{gather}
\tau_\beta\tau_\gamma\tau_\delta
+q^{2Z+1/2-2(\alpha-1/4)/3}{\tau'_\alpha\tau_0\tau_y}
\nonumber\\
\qquad {}+q^{2Z+1/2+2(\alpha-1/4)/3}{\overline\tau'_\alpha\tau_x\tau'_x}
+q^{(4Z+1)/3}\tau_\alpha\tau'_\alpha\overline\tau'_\alpha=0.
\end{gather}

Subtracting one from the r.h.s.\ of (4.14) we see that we recover the l.h.s.\ of
(4.15) up to a global multiplicative factor. Multiplying by the same factor
the r.h.s.\ of (4.15) and using the value of $K$   we find that the quantity
$(xyq^{-2Z-1/2}-1)$  becomes:
\begin{gather}
{xyq^{-2Z-1/2}-1}=
-q^{-(8Z+2)/3}\frac{\tau_\alpha\tau_\beta\tau_\gamma\tau_\delta}
{\tau_x\tau'_x\tau_0\tau_y}
\end{gather}
and similarly
between $x$ and ${\underline y}$ we have
\begin{gather}
{x{\underline y} q^{-2Z+1/2}-1}=
-q^{-(8Z-2)/3}\frac{\tau'_\alpha\tau'_\beta\tau'_\gamma\tau'_\delta}
{\tau_x\tau'_x\tau_0\tau_y}.
\end{gather}

Taking the
product, we recover the products of $\tau$'s involving the quantities
$(x-c_i)$. When we take into account
carefully all prefactors we find:
\begin{gather}
\left(xyq^{-2Z-1/2}-1\right)\left(x{\underline y}
q^{-2Z+1/2}-1\right)\nonumber\\
\qquad {}=q^{-8Z}\frac{\left(x-q^{Z+\alpha}\right)\left(x-q^{Z+\beta}\right)
\left(x-q^{Z+\gamma}\right)
\left(x-q^{Z+\delta}\right)}{\left(x-q^{-2Z+p}\right)\left(x-q^{-2Z-p}\right)}
\end{gather}
which
is one of the two equations of the system
defining the asymmetric $q$-P$_{\rm V}$ equation~[14],  although not
in its usual form. Again, in order to obtain the
second equation, we need to consider the point~$\overline x$.
Translating (4.18) by a full NV forwards, we get:
\begin{gather}
\overline  xyq^{-2Z-3/2}-1=
-q^{-(8Z+6)/3}\frac{\overline\tau'_\alpha\overline\tau'_\beta\overline\tau'_\gamma
\overline\tau'_\delta}{\overline\tau_x\overline\tau'_x\tau_0\tau_y}.
\end{gather}

Multiplying with (4.17) we find
products of $\tau$'s near $y$, and get  the analogue of (4.19):
\begin{gather}
\left(xyq^{-2Z-1/2}-1\right)
\left(\overline xyq^{-2Z-3/2}-1\right)\\
\qquad {}=q^{-8Z-4}
\frac{\left(y-q^{Z+1/2-\alpha}\right)
\left(y-q^{Z+1/2-\beta}\right)\left(y-q^{Z+1/2-\gamma}\right)
\left(y-q^{Z+1/2-\delta}\right)}{\left(y-q^{-2Z-1+r}\right)
\left(y-q^{-2Z-1-r}\right)}.\nonumber
\end{gather}
The equations (4.21) and (4.19) together
form the asymmetric  $q$-P$_{\rm V}$ equation. To recover the usual
form, we absorb $q^{-Z}$ into a redefinition of
$x$ (and appropriate factors for the other variables) to get:
\begin{subequations}
\begin{gather}
(xy-1)(x{\underline y} -1)=\frac{(x-q^{\alpha})(x-q^{\beta})(x-q^{\gamma})
(x-q^{\delta})}{(1-xq^{3Z-p})(1-xq^{3Z+p})},\\
(xy-1)(\overline xy-1)= \frac{(y-q^{-\alpha})(y-q^{-\beta})(y-q^{-\gamma})
(y-q^{-\delta})}{(1-yq^{3Z+3/2-r})(1-yq^{3Z+3/2+r})}
\end{gather}
\end{subequations}
which is the  asymmetric  $q$-P$_{\rm V}$
equation in its usual form up to a redefinition of $Z$.

The continuous limits of asymmetric  $q$-P$_{\rm V}$ and asymmetric
d-P$_{\rm IV}$ were presented in~[14],
where it was shown that both equations have P$_{\rm VI}$ as continuous limit.

\section{The contiguity relations and the Miura's}

Among the 15 NV's around $X$ which allow to reach sites of nonlinear
variables, 8 have scalar product 1/2
with the NV $(2;0,0,0,0,0)$ along $XY$ (corresponding to taking one of
the two NNNV with $a=-2$ and one of the four
ones with $a=1$) and 6 with scalar product $-1$ with this NV (taking
2 among the 4 NNNV's with $a=1$). Note that the
latter six come by pairs, the sum of two NV's of one pair being the
opposite of the NV along $XY$. This is the case,
for instance, for the NV $(-1;0,0,1,0,0)$ related to
$\alpha$ and $\beta$, say -- half the opposite of the sum of the two
NNNV's $(1;0,0,-1,-1,-1)$ and $(1;0,0,-1,1,1)$ --
which lead to the two first $C_i$'s in (4.2) and the one related to
$\gamma$ and $\delta$, namely
$(-1;0,0,-1,0,0)$. The point $\wdt_{\alpha,\beta}$
($-1/2;1/2,-1/2,-1/2,0,0)$, reached by
translating
$X$ by half the {\it opposite} of the first forms an equilateral
triangle with $X$ and~$Y$.   Similarly,
$W_{\alpha,\beta}$ at $(-3/2;1/2,-1/2,1/2,0,0)$
forms an equilateral triangle with~$X$
 and~${\underline Y}$. Contrary to the case of E$_7^{(1)}$ we considered in~[15],
where points analogous to these two points were
the only ones near
$X$ in the two-dimensional plane containing them together with $X$
and~$Y$, ${\overline Y}$, here the two points
$W_{\gamma,\delta}$ at
$(-3/2;1/2,-1/2,-1/2,0,0)$ and
$\wdt_{\gamma,\delta}$ at
$(-1/2;1/2,-1/2,1/2,0,0)$ are also in the same
plane and form, with
$Y$, ${\underline Y}$, $W_{\alpha,\beta}$ and $\wdt_{\alpha,\beta}$, a regular
hexagon of center
$X$. Note that the two $\tau$'s in NNV relative position around
$\wdt_{\alpha,\beta}$ are
$\tau_\alpha$  at
$(-1/2;1/2,-1/2,-1/2,-1/2,-1/2)$ and a
$\tau$ at point
$(-1/2;1/2,-1/2,-1/2,1/2,1/2)$ which is
what we implicitly called
$\tau_\beta$.   We will not consider all six possible pairs of NNNN
$\tau$'s around $\wdt_{\alpha,\beta}$, but two of
such pairs are precisely $\{\overline\tau'_\gamma,\tau'_\delta\}$ and
$\{\overline\tau'_\delta,\tau'_\gamma\}$,
associated to the NNNV $(-2;0,0,0,-1,1)$ and $(-2;0,0,0,1,-1)$
respectively. (We are not giving the coordinates of
all these $\tau$'s; they can be deduced from those of index $\alpha$
by changing the sign of two of the last three
components, the component which does not change sign being the
fourth, fifth and sixth one, respectively, for $\beta$,
$\gamma$ and~$\delta$). Moreover note that the $\tau_0$ at the origin
and $\tau_y$ each belong to one NNNN pair,
around $\wdt_{\alpha,\beta}$, associated to the NNNV's
$(1;-1,1,1,0,0)$ and $(1;1,-1,1,0,0)$. When $\tau'_\gamma$, say,
vanishes, both~$X$ and
$\wdt_{\alpha,\beta}$  take specific values, namely ($Z+\gamma$) and
($-2Z+(\delta-\gamma-1)/2$), respectively.  So
the sum ($X+\wdt_{\alpha,\beta}$)  takes the value
$-(Z+(\alpha+\beta+1)/2)$ (we have used the fact that the
$\alpha+\beta+\gamma+\delta=0$), which, as it turns out, is
the same when either
$\tau'_\delta$, $\tau_0$ or $\tau_y$ vanish. In fact these four
$\tau$'s play, for the pair of points
$\{X,\wdt_{\alpha,\beta}\}$ the same role as~$\tau_\alpha$, etc., for
$\{X,Y\}$. If we were considering the
asymmetric  $q$-P$_{\rm V}$ equation, the relevant nonlinear variables
$x$, $\wmdt_{\alpha,\beta}$ are just $q$ raised at a power equal to
the relevant quantities, and 
their product now takes the same 
values whether $\tau'_{\gamma}$, $\tau'_{\delta}$, $\tau'_0$ or 
$\tau'_y$ vanish.
What happens is that we have another instance of
a~trilinear equation like (4.5) (resp.~4.16).

Considering first the asymmetric  d-P$_{\rm IV}$ equation, and
following the same line of reasoning we find that:
\begin{equation}
X+\wdt_{\alpha,\beta}+Z+(\alpha+\beta+1)/2=-\frac{\tau'_\gamma
\tau'_\delta \tau_0\tau_y}{\tau_\alpha\tau_\beta\tau_x\tau'_x}.
\end{equation}
 One could easily
obtain an equation relating
$\wdt_{\alpha,\beta}$, $X$ and $W_{\alpha,\beta}$, analogous to
(4.10). But we are here interested in the Miura
relating
$\wdt_{\alpha,\beta}$, $X$ and $Y$. Let us multiply (5.1) and~(4.6).
Four $\tau$'s drop out and we are left with:
\begin{equation}
(X+Y-2Z-1/2)(X+\wdt_{\alpha,\beta}+Z+(\alpha+\beta+1)/2)=
\frac{\tau_\gamma\tau_\delta\tau'_\gamma\tau'_\delta}
{\tau_x^2{\tau'_x}{\!}^2}
\end{equation}
which, if we compute $X$ in two different ways from  (3.1) with the appropriate
instances (namely the two last ones) of  (4.2) is just:
\begin{gather}
(X+Y-2Z-1/2)(X+\wdt_{\alpha,\beta}+Z+(\alpha+\beta+1)/2)\nonumber\\
\qquad {}=(X-Z-\gamma)(X-Z-\delta).
\end{gather}

The quantity
$-(Z+(\alpha+\beta+1)/2)$ is the scalar product of the NV
$(-1;0,0,1,0,0)$ from $\wdt_{\alpha,\beta}$ to $X$, with
the vector joining
$O'$ to the midpoint of $\wdt_{\alpha,\beta} X$ and plays exactly the
same role as $2Z+1/2$  between $X$ and $Y$. We
could get an analogous equation for any of the six $\wdt$ that form
an equilateral triangle with $X$ and $Y$, with any
two among
$\alpha$,
$\beta$, $\gamma$, $\delta$ as indices of $\wdt$ and the other two
appearing in the rigth hand side. The factor
involving $X$ and $Y$ is the same for all six choices.  Equation
(5.3) on the equilateral triangle
$X$, $Y$, $\wdt_{\alpha,\beta}$ seems to singularize $X$ but this is not
true.   If we expand this equation we find:
\begin{gather}
XY+X\wdt_{\alpha,\beta}+Y\wdt_{\alpha,\beta}-(2Z+1/2)\wdt_{\alpha,\beta}\nonumber\\
\qquad {}+(Z+(\alpha+\beta+1)/2)Y+(Z-(\alpha+\beta)/2)X
=(2Z+1/2)^2/2\nonumber\\
\qquad {}+(Z+(\alpha+\beta+1)/2)^2/2+(Z-(\alpha+\beta)/2)^2/2-(\gamma-\delta)^2/4.
\end{gather}

The quadratic terms are obviously symmetric, and the coefficients of
the three linear terms are just the opposite
of the scalar product of the vector joining $O'$ to the relevant
point with the NV connecting the two others. As for
the right hand side, it is a quantity which
treats these three NV's in the same way. This is the Miura written in
a symmetric way. The translations of $X$, $Y$
that were used to  simplify (4.10) are not appropriate, but there is
still a way to simplify (5.4), by subtracting
from  each variable its coefficient in this equation. In the new
variables, one finds a very elegant result:
\begin{equation}
{\cal XY}+{\cal X}\kwdt_{\alpha,\beta}+{\cal Y}\kwdt_{\alpha,\beta}
+(\gamma-\delta)^2/4=0.
\end{equation}
Unfortunately ${\cal X}$ and ${\cal Y}$ are not exactly the same
variables as the $X$ and $Y$ of (4.13).
Equation (5.4) is the contiguity relation on a triangle, the Miura
transformation that allows to determine any
nonlinear variable in the lattice from two inital data, for the
asymmetric  d-P$_{\rm IV}$ equation.

The case of the asymmetric  $q$-P$_{\rm V}$ equation is similar. With
the same reasoning we get the relation between
$x$, $y$ and $\wmdt_{\alpha,\beta}$, first written in a way that
seems to singularize~$x$:
\begin{equation}
\left(xyq^{-2Z-1/2}-1\right)\left(x\wmdt_{\alpha,\beta}q^{Z+(\alpha+\beta+1)/2}-1\right)=
q^{2(-2Z+\alpha+\beta)/3}
\frac{\tau_\gamma\tau_\delta\tau'_\gamma\tau'_\delta}{\tau_x^2{\tau'_x}{\!}^2}
\end{equation}
and thus
\begin{equation}
\left(xyq^{-2Z-1/2}-1\right)\left(x\wmdt_{\alpha,\beta}q^{Z+(\alpha+\beta+1)/2}-1\right)=
\left(q^{-Z-\gamma}x-1\right)\left(q^{-Z-\delta}x-1\right).
\end{equation}
 Again, six similar
equations can be obtained by permuting the four
quantities $\alpha$, etc.\ and~(5.6) can also be written in a way that
is symmetric in terms of $x$, $y$ and
$\wmdt_{\alpha,\beta}$. After some elementary algebra we find:
\begin{gather}
xy\wmdt_{\alpha,\beta}=q^{-Z+(\alpha+\beta)/2}x+q^{-Z-(\alpha+\beta+1)/2}y
\nonumber\\
\qquad {}+q^{2Z+1/2}\wmdt_{\alpha+\beta}-q^{(\gamma-\delta)/2}-q^{(\delta-\gamma)/2}.
\end{gather}
The coefficients of the nonlinear variables are just $q$
raised to the opposite of the coefficients that
appear in (5.4). This shows that this equation is also invariant in
the exchange of the three variables. Obviously,
a redefinition of the variables can put all their coefficients to
unity,
\begin{equation}
\mathtt{xy}\mwdt_{\alpha,\beta}
=\mathtt{x}+\mathtt{y}+\mwdt_{\alpha,\beta}
-q^{(\gamma-\delta)/2}-q^{(\delta-\gamma)/2}
\end{equation}
  but the $\mathtt{x}$ and $\mathtt{y}$ of (5.9) are not the same
variables as the $x$
and $y$ of equation~(4.22). Equation (5.8) is  the contiguity relation on a triangle,
the Miura transformation that allows to determine any nonlinear variable in the
lattice from two inital data, for the asymmetric
$q$-P$_{\rm V}$ equation.

In Section 4, we were able to write the equation between $X$, $Y$ and
${\underline Y}$ (or $x$, $y$, ${\underline y}$) without  having to go through the
Miura's. But it is interesting to show how to recover
it from the Miura's. First, note that the various $W$ obtained by
translating $X$ forward by half of six of the NV's
form equilateral triangles with $X$ and
${\underline Y}$. In particular, we could choose $W_{\gamma,\delta}$.  Then we
have the analogues of (5.3) and (5.7) for
asymmetric  d-P$_{\rm IV}$ and  asymmetric  $q$-P$_{\rm V}$ respectively.
\begin{gather}
(X+{\underline Y}-2Z+1/2)(X+W_{\gamma,\delta}+Z+(\gamma+\delta-1)/2)\nonumber\\
\qquad {}=(X-Z-\alpha)(X-Z-\beta),\\
\left(x{\underline y} q^{-2Z+1/2}-1\right)\left(xw_{\gamma,\delta}q^{Z+(\gamma+\delta-1)/2}-1\right)=
\left(q^{-Z-\alpha}x-1\right)\left(q^{-Z-\beta}x-1\right).
\end{gather}
 But, as we mentioned
above, $W_{\gamma,\delta}$ belongs to a regular
hexagon around $X$ that contains $Y$, ${\underline Y}$,
$\wdt_{\alpha,\beta}$ and $W_{\alpha,\beta}$, and in particular,
forms an equilateral triangle with $X$ and
$\wdt_{\alpha,\beta}$. The analogues of (5.3) and (5.7) can easily be
obtained. The factors in the left hand side
involving the $W$'s, $w$'s are the {\sl same} as those in the
previously written equations, coupling them to $Y$,
$y$: indeed, they only depend on the NV relating the point under
consideration and $X$. Only the right hand side
depends on what we are coupling them to. Carefully checking which are
the $\tau$'s that end up in the right hand
side, we finally find:
\begin{gather}
(X+\wdt_{\alpha,\beta}+Z+(\alpha+\beta+1)/2)(X+W_{\gamma,\delta}+Z+(\gamma+\delta-1)/2)\nonumber\\
\qquad {}=(X+2Z-p)(X+2Z+p),\\
\left(x\wmdt_{\alpha,\beta}q^{Z+(\alpha+\beta+1)/2}-1\right)
\left(xw_{\gamma,\delta}q^{Z+(\gamma+\delta-1)/2}-1\right)
\nonumber\\
\qquad {}=\left(q^{2Z-p}x-1\right)\left(q^{2Z+p}x-1\right).
\end{gather}
   From (5.3), (5.10), (5.12) and (5.7), (5.11), (5.13) respectively, one can
easily recover (4.10) and (4.19).

\section{Conclusion}

In this paper we have presented the geometrical description of the
discrete Painlev\'e equations known as asymmetric
$q$-P$_{\rm V}$ and asymmetric {\rm d}-P$_{\rm IV}$. (Despite these
names both equations are discrete analogues of
P$_{\rm VI}$ as can be assessed through their continuous limits).
This geometrical description was performed in
the framework of the affine Weyl group~E$_6^{(1)}$. It was shown that
both discrete ${\mathbb P}$'s possess the property
of self-duality i.e.\ the same equation governs the evolution along
the individual variable or along the parameters of
the d-${\mathbb P}$ induced by the Schlesinger transformations. This
geometrical approach allows to describe all the known
d-${\mathbb P}$'s in a unified approach. Moreover it makes possible the
investigation of all possible equations related to
the basic ones (here asymmetric $q$-P$_{\rm V}$ and asymmetric {\rm
d}-P$_{\rm IV}$) by considering various evolution
paths within the geometry of E$_6^{(1)}$ (a~question we intend to
return to in some future work).

\label{Grammaticos-lastpage}

\end{document}